\documentclass[hidelinks,conference,a4paper]{IEEEtran}
\pdfoutput=1
\usepackage{textcomp}
\usepackage{xcolor}
\usepackage{footnotebackref}
\usepackage{footmisc} 
\usepackage{mathtools}
\usepackage{multicol}
\usepackage{cuted}
\usepackage{lipsum, color}
\usepackage{tabularx,booktabs}
\usepackage{hhline}
\usepackage{colortbl}
\usepackage{color}
\usepackage{amssymb}   

\usepackage{balance}            
\usepackage{bigstrut}
\usepackage{cite}               
\usepackage{threeparttable}     
\newcolumntype{M}[1]{>{\centering\arraybackslash}m{#1}}  
\usepackage{color} 
\newcommand{\Ax}{$\times$}  

\usepackage{pgffor}

\newcommand{\gemvName}{IMAGine}

\newcommand{\gemvElabUl}{\underline{I}n-\underline{M}emory \underline{A}ccelerated \underline{G}EMV eng\underline{ine}}

\IEEEoverridecommandlockouts        

\begin{document}

\bstctlcite{IEEEexample:BSTcontrol}

\title{IMAGine: An \textbf{\textit{I}}n-\textbf{\textit{M}}emory \textbf{\textit{A}}ccelerated \\ \textbf{\textit{G}}EMV Eng\textbf{\textit{ine}} Overlay}


{
\author{

\IEEEauthorblockN{
    MD Arafat Kabir\IEEEauthorrefmark{1}, 
    Tendayi Kamucheka\IEEEauthorrefmark{1}, 
    Nathaniel Fredricks\IEEEauthorrefmark{1},\\
    Joel Mandebi\IEEEauthorrefmark{2},
    Jason Bakos\IEEEauthorrefmark{3},
    Miaoqing Huang\IEEEauthorrefmark{1}, and
    David Andrews\IEEEauthorrefmark{1}
}
\IEEEauthorblockA{
\IEEEauthorrefmark{1}%
    Department of Electrical Engineering and Computer Science,
    University of Arkansas, \\ 
\IEEEauthorrefmark{3}%
    Department of Computer Science and Engineering,
    University of South Carolina, \\
\IEEEauthorrefmark{2}%
    Advanced Micro Devices, Inc. (AMD)
}

\{makabir, tfkamuch, njfredri, mqhuang, dandrews\}@uark.edu, 
jmandebi@amd.com, jbakos@cse.sc.edu,%
\thanks{This material is based upon work supported by the National Science Foundation under Grant No. 1955820.}
}   
}   


\onecolumn
{\large\vspace*{\fill}

© 2024 IEEE.  Personal use of this material is permitted.  
Permission from IEEE must be obtained for all other uses, 
in any current or future media, including reprinting/republishing this 
material for advertising or promotional purposes, 
creating new collective works, for resale or redistribution to servers or lists, 
or reuse of any copyrighted component of this work in other works. \\

This work has been accepted at the 
2024 34th International Conference on Field-Programmable Logic and Applications (FPL)
and will appear in the proceedings and on the IEEE website soon.

\vspace*{\fill}
}
\twocolumn

\maketitle

\begin{abstract}
Processor-in-Memory (PIM) overlays and alternative reconfigurable tile fabrics have been proposed
to eliminate the von Neumann bottleneck and enable processing performance to scale with BRAM capacity. 
The performance of these FPGA-based PIM architectures has been limited due to a reduction of the BRAMs maximum 
clock frequencies and less than ideal scaling of processing elements with increased BRAM capacity. 
This paper presents {\gemvName}, an {\gemvElabUl}, a PIM-array accelerator that
clocks at the maximum frequency of the BRAM and scales to 100\% of the available BRAMs.
Comparative analyses are presented showing execution speeds over existing PIM-based GEMV engines on FPGAs and
achieving a \mbox{2.65\Ax\ -- 3.2\Ax} faster clock.
An AMD Alveo U55 implementation is presented that achieves a system clock speed of 737 MHz, 
providing 64K bit-serial multiply-accumulate (MAC) units for GEMV operation.
This establishes {\gemvName} as the fastest PIM-based GEMV overlay, outperforming even 
the custom PIM-based FPGA accelerators reported to date.
Additionally, it surpasses TPU v1-v2 and Alibaba Hanguang 800 in clock 
speed while offering an equal or greater number of multiply-accumulate (MAC) units.
\end{abstract}


\section{Introduction}
The exponential growth of Internet-of-Things (IoT) devices and social media applications 
has significantly changed the landscape of computing workloads.
Modern workloads, such as scientific computation, graph processing, and machine learning, generate and process datasets that
are expanding at a rate that outpaces Moore's Law\cite{tpuv4_2021}.
However, today's processors remain constrained by the ``Memory Wall'' of the von Neumann architecture,
which limits the ability to exploit the parallelism within these memory-intensive tasks. 
Processing-in-memory (PIM) architectures are being 
pursued~\cite{zpim2020,pimca2023,sramcim2022,skhynixPim_2023,ccb2021,spar2020,spar2021,spar2thesis_2022,comefa2022,comefa2023,bramac2023,m4bram2023,picasoposter2023,picaso2023}
to mitigate the memory wall and enable processing performance to scale with memory capacity.

Modern Field Programmable Gate Arrays (FPGAs) with 100s of Mbits of SRAM distributed throughout the device 
in the form of disaggregated memory resources can provide several TB/s of internal bandwidth.  This is an ideal programmable substrate for creating customized Processor In/Near Memory accelerators. 
Several PIM array-based accelerator designs~\cite{ccb2021,spar2020,spar2021,spar2thesis_2022,comefa2022,comefa2023,bramac2023,m4bram2023}
have been proposed to harness this massive internal bandwidth.
However, results reported to date show 
achievable clock frequencies
and compute densities are not sufficient to compete with their custom 
Application Specific Integrated Circuit (ASIC) counterparts.

Such shortcomings have motivated 
redesigns of the separate Block-RAM (BRAM) and LUT resources into tightly integrated PIM tiles.  
While these redesigns have increased chip compute densities, 
the maximum achievable clock frequency remains no better than their overlay counterparts.  
Additionally, the adoption of a bigger FPGA with an increased resource capacity does not translate into a linear increase in compute performance.

This paper presents {\gemvName}, a PIM array-based GEMV accelerator that clocks at the maximum frequency of the BRAM. 
The PIM tile array architecture of {\gemvName} has been designed to achieve linear scalability of the number 
of compute units with increased BRAM densities. 
Comparative studies are presented that show it is the fastest and most 
scalable PIM array-based GEMV accelerator reported to date.
Run time results also show that {\gemvName} shatters some of the myths concerning performance 
limitations of PIM-array accelerators and FPGA overlays in general.
Our contributions can be summarized as follows,
\begin{itemize}
    \item A set of aspirational but practical design goals for PIM array-based accelerators.
          We argue these goals need to be met to claim a ``Scalable High-Performance PIM design'' on FPGAs.
    \item We present the design and implementation of {\gemvName}, an {\gemvElabUl} overlay, 
          that breaks some existing myths around FPGA design, clocking faster than Google's TPU v1-v2
          with equal or more processing elements (PEs) using an off-the-shelf datacenter-grade FPGA. 
    \item We present a comparative study of {\gemvName} with existing PIM-array accelerators, 
          establishing it as the fastest and most scalable FPGA PIM-based GEMV accelerator.
\end{itemize}

IMAGine has been published at~\cite{github_imagine} as open-source 
implementation and is freely available for study, use, modification, 
and distribution without restriction.

\section {Related Work}
\label{sec:background}

\subsection{Custom-BRAM PIMs}
Wang et al~\cite{ccb2021} proposed the Compute-Capable BRAM (CCB) based on Neural Cache~\cite{neuralcache2018}. 
CCB exposes compute parallelism within a BRAM by converting each BRAM bitline into a bit-serial Processing Element (PE).
CCB was used to build RIMA~\cite{ccb2021} to accelerate recurrent neural networks (RNNs).
RIMA achieved 1.25\Ax\ and 3\Ax\ higher performance compared to 
the Brainwave DL soft processor\cite{msbrainwaveNPU_2018} for 8-bit integer and block floating-point precisions, respectively.

\begin{table}
\setlength{\tabcolsep}{2.0pt}
\caption{Maximum Frequency (MHz) of Existing FPGA-PIM Designs}
\label{tab:pim-speeds}
\centering
\begin{tabular}{c|ccc|cc|cc}
\hline
PIM Design & Type   & Device & \textsl{f$_{BRAM}$}   & \textsl{f$_{PIM}$}    & Rel.   & \textsl{f$_{Sys}$}  & Rel. \bigstrut\\
\hline
CCB        & Custom  & Stratix 10  & 1000   & 624    & 62\%   & 455    & 46\% \bigstrut[t]\\
CoMeFa-A   & Custom  & Arria 10    & 730    & 294    & 40\%   & 288    & 39\%   \\
CoMeFa-D   & Custom  & Arria 10    & 730    & 588    & 81\%   & 292    & 40\%   \\
BRAMAC-2SA & Custom  & Arria 10    & 730    & 586    & 80\%   & -      & -      \\
BRAMAC-1DA & Custom  & Arria 10    & 730    & 500    & 68\%   & -      & -      \\
M4BRAM     & Custom  & Arria 10    & 730    & 553    & 76\%   & -      & -      \\
SPAR-2     & Overlay & UltraScale+ & 737    & 445    & 60\%   & 200    & 27\%   \\
PiCaSO     & Overlay & UltraScale+ & 737    & 737    & 100\%  & -      & -      \\
\hline
\end{tabular}%
\end{table}

Arora et al~\cite{comefa2022,comefa2023} proposed CoMeFa
that uses bit-serial PEs per SRAM bitline like CCB, but exploits 
the dual-port nature of BRAMs to simultaneously read two operands. 
To evaluate the performance and energy benefits of CoMeFa RAMs, various microbenchmarks, including 
General Matrix-Vector Multiplication (GEMV) and General Matrix-Matrix Multiplication (GEMM) were studied in~\cite{comefa2023}.
Augmenting an Intel Arria 10-like FPGA with CoMeFa RAMs delivered a geomean speedup of 2.55× across diverse applications.

Chen et al proposed BRAMAC~\cite{bramac2023} and M4BRAM~\cite{m4bram2023},
which bypass MAC computation on the slow and power-hungry primary BRAM array 
by copying operands to a smaller ``dummy array".
BRAMAC requires 2-/4-/8-bit predefined weights and activations,
limiting its use to quantized uniform-precision deep neural nets. 
M4BRAM overcomes some of these limitations by enabling variable activation
precision between 2 and 8 bits with linearly scaled MAC latency.
Combining BRAMAC-2SA/BRAMAC-1DA with Intel's DLA~\cite{intelDLA_2017} resulted in an average speedup 
of 2.05×/1.7× for AlexNet and 1.33×/1.52× for ResNet-34. 
M4BRAM surpassed BRAMAC by an average of 1.43× across diverse benchmarks.

\subsection{BRAM-Overlay PIMs}
To leverage the benefits of PIM architectures in contemporary FPGAs, PIM overlay architectures have been proposed.
Panahi et al~\cite{spar2020, spar2021, spar2thesis_2022} proposed SPAR-2, a SIMD PIM-array overlay accelerator,
connecting bit-serial PEs from the programmable fabric with BRAMs.
SPAR-2 was implemented on Virtex-7 and Virtex UltraScale FPGAs with 10K PEs to accelerate
several deep learning applications.
It achieved up to 34.2× and 3.5× speedups compared to other custom HLS-based and RTL-based accelerators, respectively.

Building upon the PIM overlay of SPAR-2, Kabir et al proposed PiCaSO~\cite{picaso2023}
with configurable pipeline stages along the datapath. 
PiCaSO introduced an intermediate muxing module to enable zero-copy in-block reduction
and a ``binary-hopping'' pipelined NEWS network for array-level reduction.
PiCaSO provided competitive performance and memory utilization 
efficiency compared to both CCB and CoMeFa custom-BRAM architectures.

\section{Motivation and Design Goals}
\label{sec:standard}

Table~\ref{tab:pim-speeds} summarizes the maximum frequencies of the PIM designs discussed in section \ref{sec:background}.
The relative frequency columns (Rel.) show that the clock frequency \textsl{f$_{PIM}$} of all the PIM tiles
are significantly slower compared to the maximum frequency for the device BRAMs (\textsl{f$_{BRAM}$}), except for PiCaSO.
Their fastest system frequencies (\textsl{f$_{Sys}$}) are 2.1\Ax\ -- 3.7\Ax\ slower than the BRAM maximum frequencies (\textsl{f$_{BRAM}$}).
This slower frequency was attributed to the limitations of the soft logic and the routing resources of the FPGAs.
It was also reported as unlikely that an FPGA accelerator
at the system level would operate at a frequency surpassing the degraded frequency (\textsl{f$_{PIM}$}) of these PIM designs,
even in a more advanced node than the evaluation platforms~\cite{comefa2022,comefa2023,m4bram2023,bramac2023}. 

Further observation yielded that most of these systems could not utilize all available BRAMs as PIMs.
This lower utilization combined with a lower clock frequency results in less efficient use of 
the available internal BRAM bandwidth of the devices and a lower system-level compute density. 
A final observation shows a troubling common pattern: as the utilization of BRAMs
increases the achievable system-level clock frequency decreases~\cite{ccb2021,comefa2023}.

\begin{table}
\centering
\setlength{\tabcolsep}{3.5pt}
\caption{Delay (ns) Breakdown of 1-level Logic Path in AMD Devices}
\label{tab:celldelay}
\begin{threeparttable}
\begin{tabular}{cccc|c|ccc}
\hline
       & FF-C2Q\tnote{1} & LUT     & FF-Setup & Total\tnote{2}  & BRAM\tnote{3} & Net Budget & SB-Min\tnote{4} \bigstrut\\
\hline
V7     & 0.290   & 0.34   & 0.255  & 0.885  & 1.839   & 0.954 & 0.272 \bigstrut[t]\\
US+    & 0.087   & 0.15   & 0.098  & 0.335  & 1.356   & 1.021 & 0.102 \bigstrut[b]\\
\hline
\end{tabular}%
\begin{tablenotes}
\item[1] Clock-to-Q delay of flip-flops
\item[2] Total cell delay
\item[3] BRAM pulse-width requirement, clock period for Fmax
\item[4] Minimum net delay through a switchbox
\end{tablenotes}
\end{threeparttable}
\end{table}

\begin{figure}
\centering
\includegraphics[width=\linewidth]{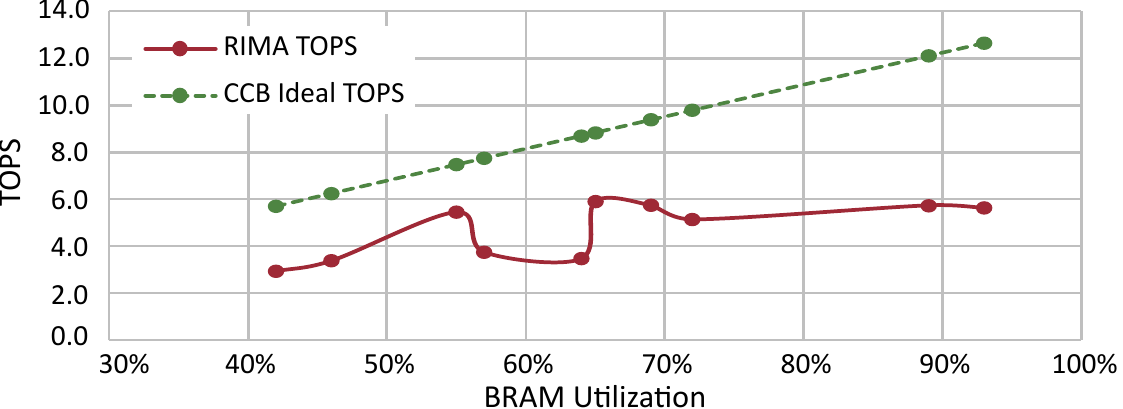}
\caption{Ideal scaling vs. actual TOPS of RIMA on Stratix 10 GX2800}
\label{fig:stdscaling}
\end{figure}

These observations motivated our interest in understanding if these results were a new reality 
of BRAM PIM arrays or symptomatic of specific design and implementation choices.

\subsection{System Clock Speed Goal}
\label{sec:idealClock}
In FPGAs, BRAMs are the single component with the longest latency~\cite{v7datasheet_2021, ultrascale_datasheet_2021, stratix10_datasheet}. 
Thus, we propose using the maximum frequency (Fmax) of the BRAM as the target frequency for the PIM-array accelerators.
To assess the practicality of this design goal, we closely examined two AMD FPGA families: Virtex-7 and UltraScale+. 
We created a test design where all timing paths are one logic level deep and averaged all paths to obtain Table~\ref{tab:celldelay}.
The Total column sums the cell delays in the columns to its left.
The BRAM column lists the clock period for BRAM Fmax. 
The SB-Min column displays the minimum delay of a net passing through a switchbox. 
Net Budget is derived by subtracting the Total column from the BRAM column.
Comparing the net budget with the minimum net delay shows that, it is feasible to design
at least two LUTs deep logic paths clocking at the BRAM Fmax.

\subsection{Performance Scaling Goal}
We posited that the peak-performance of a PIM design needs to scale linearly with the on-chip BRAM resource. 
The compute capacity in custom-BRAM-based PIM designs~\cite{ccb2021, comefa2022, comefa2023, bramac2023, m4bram2023} 
scales linearly with BRAM count if all BRAM tiles are used in PIM mode.
However, a significant sacrifice is imposed in the clock frequency that ends up limiting the achievable peak-performance on the device. 
Table~\ref{tab:pim-speeds}  \textsl{f$_{PIM}$} column indicates that the custom-BRAM PIM designs are up to 2.5× slower than the BRAM Fmax.
Fig.~\ref{fig:stdscaling} plots RIMA's peak-performance from Table-II of~\cite{ccb2021}, 
computed using reported BRAM utilization and M-DPE clock frequency. 
The irregular trend is attributed to RIMA's system-level architecture.
If RIMA adhered to the proposed performance scaling goal, even at the degraded CCB frequency of 624~MHz,
its peak-performance would align with the CCB Ideal TOPS line. 
The gap between these plots represents wasted compute capacity and memory bandwidth provided by CCB BRAMs.

\section{{\gemvName} Architecture}
\label{sec:casestudy}

\begin{figure}
\centering
\includegraphics[width=\linewidth]{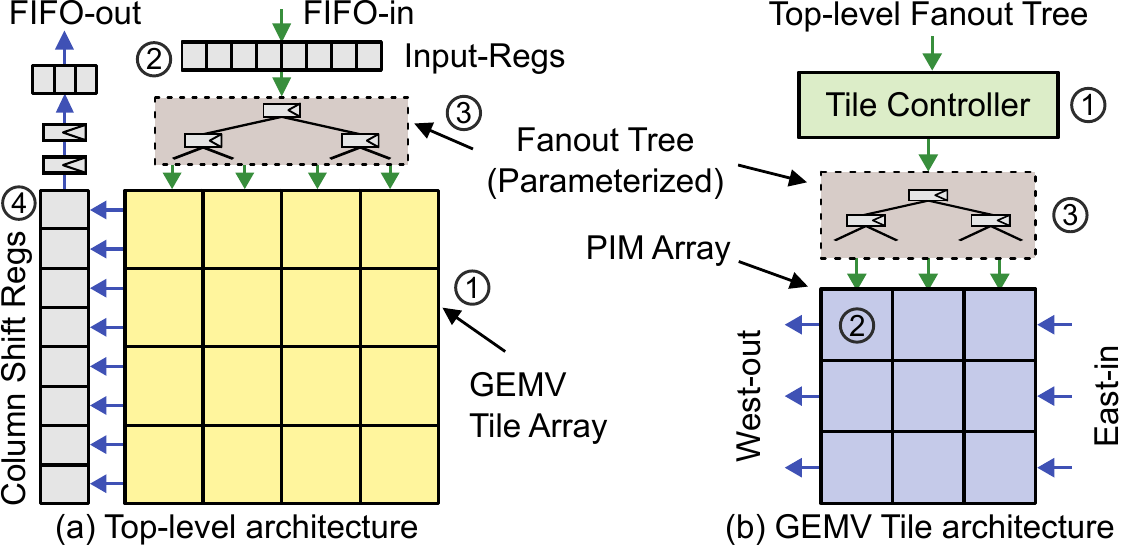}
\caption{
System architecture of {\gemvName} illustrating the data and instruction flow
(a) through the GEMV engine and
(b) within GEMV tiles.
}
\label{fig:archSysTile}
\end{figure}

\subsection{System-Level Architecture}
The top-level system is illustrated in Fig.~\ref{fig:archSysTile}(a).
It consists of (1) a 2D array of GEMV tiles, (2) a set of input registers, (3) a fanout tree connecting
the input registers to the tile array, and (4) a column of shift-registers to read out the final result.
The front-end processor sends instructions to the GEMV tiles through the input registers.
The fanout tree is parameterized to be adjusted during implementation.
The 2D tile array is implemented as a parameterized module that instantiates and connects the tiles.
At the end of the GEMV operation, the output vector is stored in the column shift registers,
which is shifted up and read through the FIFO-out port, one element per cycle.

\subsection{{\gemvName} GEMV Tile Architecture}
\label{sec:tilearch}
Illustrated in Fig.~\ref{fig:archSysTile}(b), the GEMV tile is the heart of {\gemvName}.
It consists of (1) an FSM-based controller, (2) a 2D array of PIM blocks, and (3) a fanout tree between them.
The controller receives the instruction written to the input registers at the top level,
decodes it, and generates the sequence of control signals needed to execute the instruction.
The fanout tree connects the control signals to all PEs in the PIM array
and is parameterized for adjustment during implementation. 
The PIM array interfaces allow cascading with arrays in neighboring tiles on each side. 
During accumulation, partial results move from east to west through PIM arrays, 
ultimately accumulating in the left-most PE column of the left-most GEMV tile.

\begin{figure}
\centering
\includegraphics[width=\linewidth]{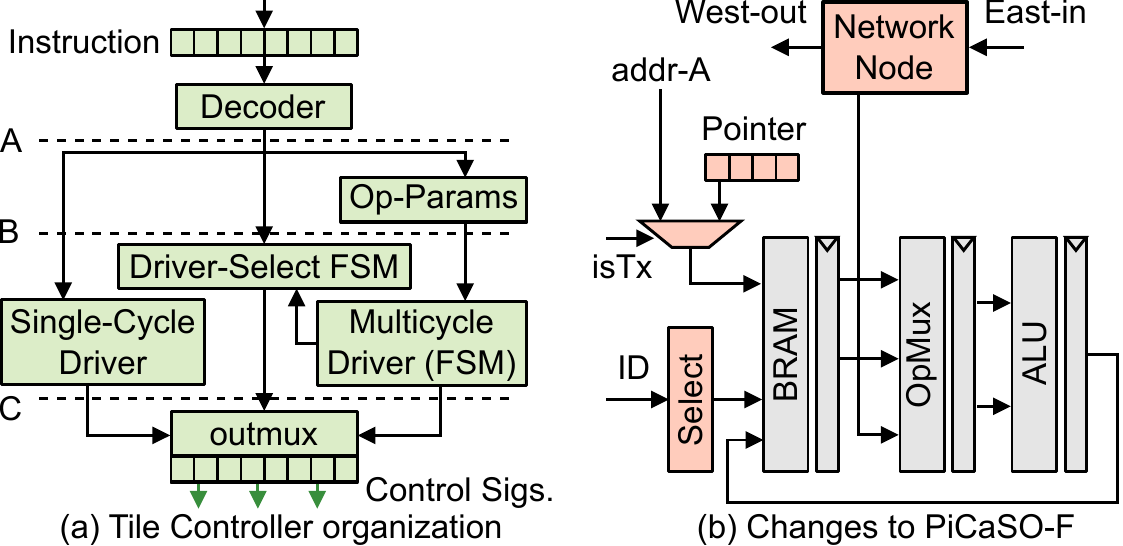}
\caption{
Architectures of 
(a) GEMV controller and 
(b) PiCaSO-IM, the adapted version of PiCaSO-F~\cite{picaso2023}.
}
\label{fig:archControlPim}
\end{figure}

\subsection{Tile Controller}
Fig.~\ref{fig:archControlPim}(a) shows the architecture of the tile controller.
It takes a 30-bit instruction, which is executed by either the single-cycle or the multicycle driver,
selected by the 2-state driver-selection FSM. 
The single-cycle driver can execute one instruction every cycle, while the multicycle driver takes several
cycles to execute instructions like ADD, SUB, MULT, etc. 
including an additional cycle to load its parameters from the Op-Params module.
All inputs and outputs are registered to localize timing paths within the controller.
The combinatorial logic in the controller is grouped into meaningful steps and optional pipeline stages
are added as illustrated by the dashed lines A, B, and C in Fig.~\ref{fig:archControlPim}(a).
Running synthesis, we ensured that each step could be implemented in one or two logic levels.

\subsection{PIM Module}
\label{sec:modpim}
We adopted PiCaSO~\cite{picaso2023} as IMAGine's PIM module for the following three reasons: 
(1) it is publicly available and open-source~\cite{github_picaso}, 
(2) it is a modifiable overlay that can be ported and studied on existing AMD devices, and 
(3) PiCaSO-F, a pipelined configuration of PiCaSO, can be clocked at the BRAM Fmax.
The modifications highlighted in red in Fig.~\ref{fig:archControlPim}(b) were
applied to PiCaSO-F to build PiCaSO-IM for IMAGine.
The original NEWS network was replaced with a simpler east-to-west data movement network.
Block-ID-based selection logic was included in PiCaSO-IM.
IMAGine's accumulation algorithm requires 3 addresses to maximize the overlap of data movement and computation.
As PiCaSO-F supports only 2 simultaneous addresses, we added a pointer register for the third address.
If PiCaSO is realized as a custom-BRAM tile as proposed in~\cite{picaso2023}, these changes can be
implemented in programmable logic fabric, keeping registerfile, OpMux, and ALU modules within the BRAM tile.
We name such a custom-BRAM implementation of PiCaSO-IM as PiCaSO-CB.

\section{Implementation and Analysis}
\label{sec:results}

In this section, we discuss the bottom-up implementation and analysis of {\gemvName},
targeting the design goals discussed in Section~\ref{sec:standard}.
In~\cite{picaso2023}, 
PiCaSO was studied on AMD Alveo U55C (xcu55c, \mbox{-2} speed grade).
We use the same device as our implementation platform to keep the results predictable.
The BRAM Fmax on this device is 737~MHz~\cite{ultrascale_datasheet_2021}, which sets the target clock period to be 1.356~ns.
All of the following studies were carried out using Vivado 2022.2.

\begin{table}
\setlength{\tabcolsep}{2.0pt}
\caption{Utilization and Frequency of 12$\times$2 GEMV Tile Components}
\label{tab:tile}
\centering
\begin{tabular}{c|cc|cc|cc|c}
\hline
       & Controller &  Rel. & Fanout &  Rel. & PIM Array &  Rel. & Tile \bigstrut\\
\hline
LUT    & 167    & 5.8\%  & 0      & 0.0\%  & 2736   & 94.2\% & 2903 \bigstrut[t]\\
FF     & 155    & 4.0\%  & 615    & 15.9\% & 3096   & 80.1\% & 3866 \\
DSP    & 0      & -      & 0      & -      & 0      & -      & 0 \\
BRAM   & 0      & 0.0\%  & 0      & 0.0\%  & 12.0   & 100.0\% & 12 \\
Freq. (MHz) & 890   & 1.2$\times$   & 890    & 1.2$\times$   & 737    & 1$\times$     & 737  \bigstrut[b]\\
\hline
\end{tabular}%
\end{table}

\subsection{GEMV Tile}
The components of the GEMV tile were studied individually to verify if they met the design requirements.
Each tile contains a 12\Ax2 PIM array and 2 stages of pipeline in the fanout tree,
which best fits the physical layout of the Alveo U55 FPGA as discussed later in this section.
Table~\ref{tab:tile} shows the utilization and performance of these components and
their relative values compared to the entire GEMV tile.

The controller together with the fanout network passed the timing constraints at a clock rate of 890 MHz.
Because the PIM array contains the BRAM, it cannot run faster than the BRAM Fmax.
It passed the timing at 737 MHz, the BRAM Fmax. 
As observed in Table~\ref{tab:tile}, the logic utilization of the controller is around 5\% of 
the entire tile and requires no DSPs, while around 90\% of the logic resources are consumed by the PIM array.
Thus, the controller and the fanout tree are not expected to bottleneck system frequency or utilization.
The GEMV tile's speed and scalability are fundamentally dependent on the PIM array, which is the desired outcome.

\subsection{Scalability Study}
\begin{figure}
\centering
\includegraphics[width=\linewidth]{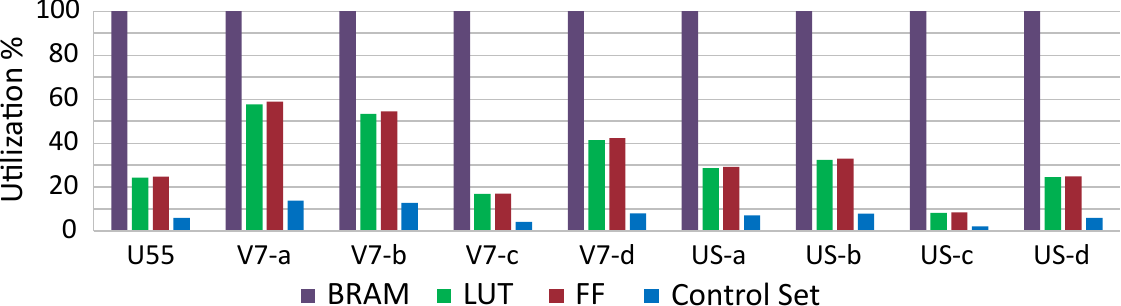}
\caption{Resource usage of {\gemvName} on representatives of Virtex-7 and Ultrascale+ families utilizing 100\% BRAMs as PIM overlays.}
\label{fig:deviceScale}
\end{figure}

To evaluate the scalability of our architecture on different device families, we followed the approach in~\cite{picaso2023}.
Along with Alveo U55, four representatives were selected from AMD's Virtex-7 and UltraScale+ devices based on two criteria: BRAM capacity and LUT-to-BRAM ratio.
Table~\ref{tab:devices} lists these devices with their BRAM capacity, LUT-to-BRAM ratio, and a short ID used in Fig.~\ref{fig:deviceScale}.
The target clock frequency of the system was set to 100 MHz on all devices to avoid timing issues 
and only focus on the logic utilization of the system at this point.

Fig.~\ref{fig:deviceScale} shows a bar graph of post-implementation utilization numbers of {\gemvName} 
on the representative devices. 
As observed, {\gemvName} can utilize 100\% of the available BRAMs as PIM overlays
providing 64K PEs in U55, with only 25\% logic and 6\% control set utilization.
This leaves sufficient logic resources to implement the fanout trees and pipeline 
stages if they are needed to achieve the target clock speed.
In fact, {\gemvName} scaled up to 100\% of available BRAM in all the representative devices for Virtex-7 and UltraScale+ families.

In the Virtex-7 family, the device V7-a has the smallest number of BRAMs and the smallest LUT-to-BRAM ratio.
{\gemvName} used around 60\% logic resources to provide 24K PEs in V7-a.
In the UltraScale+ family, US-a and US-b have the smallest number of BRAMs and the smallest LUT-to-BRAM ratio, respectively.
In these devices {\gemvName} provide 23K and 67K PEs, respectively, using roughly 30\% logic resources.
For devices with more BRAMs and a higher LUT-to-BRAM ratio the logic utilization is very small:
the logic utilization in US-c is less than 10\% providing 69K PEs.
Thus, IMAGine is scalable up to 100\% BRAM capacity irrespective of 
the available logic resources in existing devices.

\begin{table}
\centering
\begin{threeparttable}
\caption{Representatives of Virtex-7 and Ultrascale+ Families~\cite{picaso2023}}
\label{tab:devices}
\begin{tabular}{cccccc}
\hline
Device & Tech   & BRAM\# & Ratio\tnote{1} & Max PE\#\tnote{2}  & ID \bigstrut[t] \\
\hline
xcu55c-fsvh-2   & US+    & 2016   & 646    & 64K      & U55 \bigstrut[t]\\
xc7vx330tffg-2  & V7     & 750    & 272    & 24K      & V7-a \\
xc7vx485tffg-2  & V7     & 1030   & 295    & 32K      & V7-b \\
xc7v2000tfhg-2  & V7     & 1292   & 946    & 41K      & V7-c \\
xc7vx1140tflg-2 & V7     & 1880   & 379    & 60K      & V7-d \\
xcvu3p-ffvc-3   & US+    & 720    & 547    & 23K      & US-a \\
xcvu23p-vsva-3  & US+    & 2112   & 488    & 67K      & US-b \\
xcvu19p-fsvb-2  & US+    & 2160   & 1892   & 69K      & US-c \\
xcvu29p-figd-3  & US+    & 2688   & 643    & 86K      & US-d \\
\hline
\end{tabular}%
\begin{tablenotes}
\item[1] LUT-to-BRAM ratio
\item[2] Number of PEs utilizing all BRAMs as PIMs
\end{tablenotes}
\end{threeparttable}
\end{table}

\subsection{System-Level Timing Optimizations}
For the final implementation, the target clock was set to 1.356~ns to match the BRAM Fmax of Alveo U55.
The goal of the study was to find out how close we can get to the target clock rate, 
and what are the practical challenges that limit us from achieving it.
We ran the first iteration using the default settings of Vivado and achieved a setup slack of -0.52~ns.
The critical paths were within the controller with a logic depth of 4, going through the pipeline stage A
of the controller as shown in Fig.~\ref{fig:archControlPim}(a).
So, we enabled the pipeline stage A in the controller for the next iteration. 

At the end of the second iteration of implementation, we achieved a setup slack of -0.38~ns.
The control signals between the controller and the PIM array were failing 
the timing due to their high fanout and long routes.
Thus, we synthesize a fanout tree between the controller and the PIM array
empirically choosing 2 levels and a fanout of 4 for the next iteration.

\begin{figure}
\centering
\includegraphics[width=\linewidth]{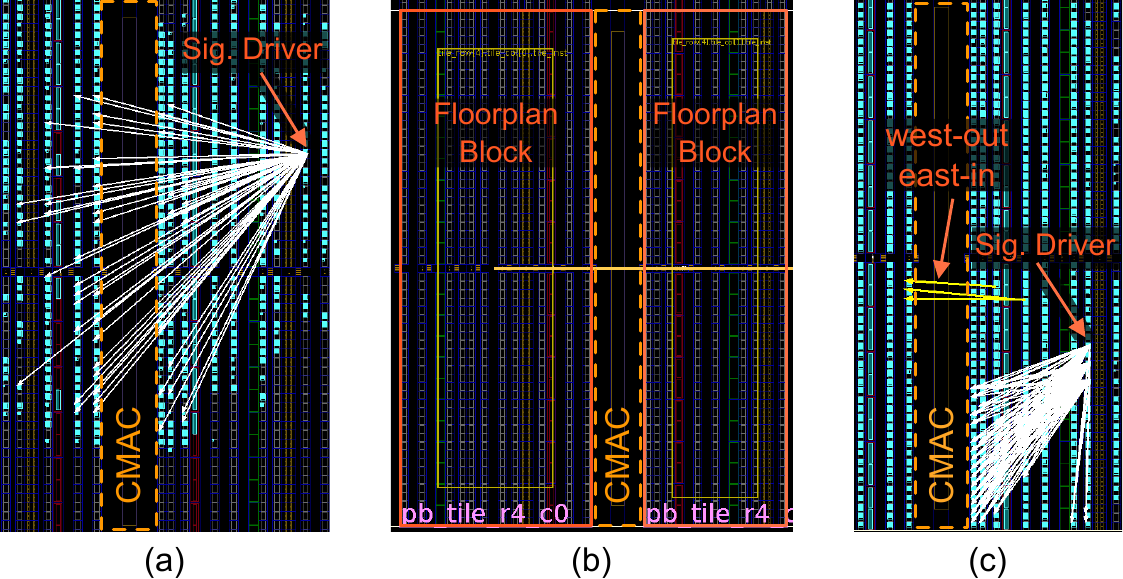}
\caption{
Avoiding unnecessary hard-block (CMAC) crossing by floorplanning
(a) placement and net connections before floorplanning,
(b) floorplan localizing logic and routing,
(b) placement and net connections in the final design.
}
\label{fig:floorplan}
\end{figure}

The design achieved a setup slack of -0.27~ns in the third iteration.
The long routes crossing hard blocks, like an Ethernet port (CMAC)~\cite{alveoU55ug_2023}, were failing the timing.
The white lines in Fig.~\ref{fig:floorplan}(a) highlight some of those critical nets. 
To avoid placement results generating such paths, we created floorplanning blocks (Pblocks)~\cite{pblockVivado_2023}
as shown in Fig.~\ref{fig:floorplan}(b), to localize the logic placement and routing of a tile. 
This required defining a tile with 12\Ax2 PIM array on Alveo U55.
Fig.~\ref{fig:floorplan}(c) shows the placement and net connections in the final iteration.
The logic and routing of each tile are localized on either side of the hard block.
Only the inter-tile connections for east-to-west accumulation, highlighted in yellow lines,
cross the CMAC block requiring minimal routing resources.

\begin{figure*}
\centering
\includegraphics[width=\linewidth]{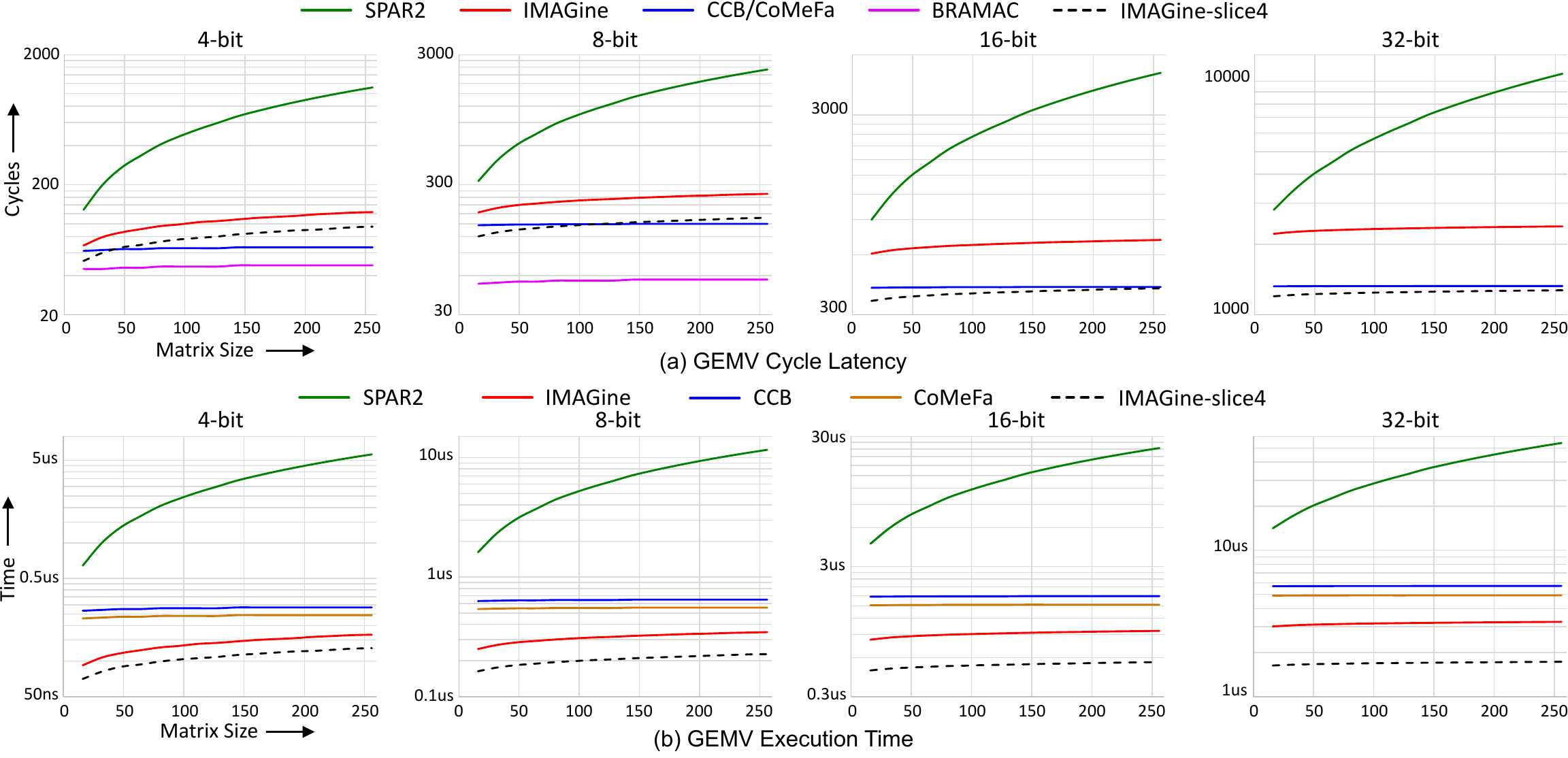}
\caption{Cycle latency and execution time of GEMV operation on different PIM array-based FPGA accelerators}
\label{fig:gemvlatency}
\end{figure*}

The final design met the timing at 737~MHz clock,
demonstrating the practical achievability of the proposed clocking goal.
Utilizing 100\% available BRAMs as PIMs, this design also achieved linear scaling of peak-performance.
Surprisingly, this clock rate is faster than custom GEMM accelerator ASICs TPU v1-v2~\cite{tpu2017, tpuv4_2021} 
and Alibaba Hanguang 800~\cite{alibabaNPU2020}, that run at 700~MHz.
Both Alveo U55 and TPU v2 are manufactured at 16~nm and Hanguang 800 at 12nm technology nodes.
So, this clock improvement is not due to a technology node difference. 
On Alveo U55, {\gemvName} has an equal number of PEs compared to TPU v1 (64K), and 4\Ax\ of TPU v2 (16K).
However, {\gemvName} can only deliver up to 0.33 TOPS at 8-bit precision, which is significantly smaller compared to 
TPU v1 (92 TOPS) and v2 (46 TOPS), due its bit-serial architecture. 
These results dispel the myth that FPGA designs are always slower and have less compute density compared to ASICs.

\subsection{Comparison With Other PIM-Array Accelerators}
Table~\ref{tab:systemcomp} shows the utilization and system frequencies of existing GEMV engines and equivalent PIM-array accelerators.
System-level utilizations and frequencies for BRAMAC and M4BRAM-based systems were not reported in~\cite{bramac2023, m4bram2023}.
RIMA~\cite{ccb2021} was evaluated on a Stratix 10 GX2800 FPGA with a BRAM Fmax of 1~GHz~\cite{stratix10_datasheet}.
Its fastest reported configuration (RIMA-Fast) runs at 455~MHz, which is 2.2\Ax\ slower than its BRAM Fmax. 
The largest reported configuration (RIMA-Large) utilizes 93\% of BRAMs and runs at 278~MHz,
4\Ax\ slower than the BRAM Fmax. 
The GEMV/GEMM systems based on CCB and CoMeFa 
were evaluated on an Arria 10 GX900 with a BRAM Fmax of 730~MHz~\cite{comefa2023}.
Though CoMeFa-based designs run slightly faster than the CCB-GEMV engine,
they are still roughly 3\Ax\ slower than the BRAM Fmax. 
Thus, CCB and CoMeFa-based GEMV/GEMM engine performance did not scale well at the system level.

SPAR-2~\cite{spar2021} utilized only 30\% of the BRAMs while running 4\Ax\ slower than BRAM Fmax on both platforms.
Thus, its performance and scalability are even worse than CCB and CoMeFa-based systems.
On the other hand, {\gemvName} has a system clock running at the BRAM Fmax while utilizing 100\% device BRAM as PIMs.
Outperforming all existing designs, {\gemvName} is the fastest PIM array-based GEMV engine implemented on any FPGA, 
running at a clock rate 2.65\Ax\ -- 3.2\Ax\ faster than any existing design.
This is an important proof of concept design that dispels earlier beliefs that 
PIM arrays and overlay accelerators cannot achieve BRAM Fmax clock frequencies at the system level~\cite{comefa2022,comefa2023,m4bram2023,bramac2023}.

As observed in Table~\ref{tab:systemcomp}, RIMA and CCB/CoMeFa-based designs 
exhaust either the logic resources or the DSPs of the device even though 
their PIM blocks are implemented by customizing the BRAM tile itself.
Even after being an overlay, {\gemvName} is achieving faster clock and better scalability using 0 DSPs and 
only one-third of the device logic resources due to its near-optimal architectural choices. 
Like SPAR-2, {\gemvName} does not use DSPs to implement the bit-serial PEs.
With a custom-BRAM implementation of the PIM module, like PiCaSO-CB discussed in Section~\ref{sec:modpim}, 
{\gemvName} would consume about 10\% of device resources while being fully scalable and implementable even in resource-limited FPGAs.

\begin{table}
\setlength{\tabcolsep}{3.5pt}
\caption{Utilization and Frequency of PIM-based GEMV/GEMM Engines}
\label{tab:systemcomp}
\centering
\begin{threeparttable}
\begin{tabular}{c|cc|cc|cc}
\hline
       & LUT    & FF     & DSP    & BRAM   & \textsl{f$_{Sys}$}\tnote{1} & Rel. Freq  \bigstrut\\
\hline
RIMA-Fast & \multicolumn{2}{c|}{60\%} & 50\%   & 55\%   & 455    & 45.5\% \bigstrut[t]\\
RIMA-Large & \multicolumn{2}{c|}{89\%} & 50\%   & 93\%   & 278    & 27.8\% \\
CCB GEMV & \multicolumn{2}{c|}{27.9\%} & 90.1\% & 91.8\% & 231    & 31.6\% \\
CoMeFa-A GEMV & \multicolumn{2}{c|}{27.9\%} & 90.1\% & 91.8\% & 242    & 33.2\% \\
CoMeFa-D GEMM & \multicolumn{2}{c|}{25.5\%} & 92.4\% & 86.7\% & 267    & 36.6\% \\
SPAR-2 (US+) & 11.3\% & 2.4\%  & 0.0\%  & 14.5\% & 200    & 27.1\% \\
SPAR-2 (V7) & 28.5\% & 7.0\%  & 0.0\%  & 30.4\% & 130    & 23.9\% \\
{\gemvName} & 35.6\% & 24.8\% & 0.0\%  & 100.0\% & 737    & 100.0\% \\
{\gemvName}-CB\tnote{2} & 10.1\% & 7.2\%  & 0.0\%  & 100.0\% & 737    & 100.0\% \bigstrut[b]\\
\hline
\end{tabular}%
\begin{tablenotes}
\item[1] System frequency in MHz
\item[2] {\gemvName} with custom-BRAM PIM tile (PiCaSO-CB)
\end{tablenotes}
\end{threeparttable}
\end{table}

\subsection{GEMV Execution Latency}
Fig.~\ref{fig:gemvlatency} plots the GEMV latency of PIM-array accelerators, 
with square-matrix dimensions on the x-axis and latency in log scale on the y-axis.
The execution times in Fig.~\ref{fig:gemvlatency}(b) are computed by multiplying cycle latencies 
with the corresponding clock periods from Table~\ref{tab:systemcomp} system frequencies.
We adopted the approach in~\cite{bramac2023} to model the block-level cycle latencies of
CCB, CoMeFa, BRAMAC, and SPAR-2 using their analytical models.
{\gemvName}'s latency model was developed and validated by running a prototype on hardware.

As observed in Fig.~\ref{fig:gemvlatency}(a), BRAMAC has the shortest cycle latency,
due to their hybrid bit-serial \& bit-parallel MAC2 algorithm.
BRAMAC's MAC latency grows linearly with operand bit-width,
while it grows quadratically in the other bit-serial architectures. 
BRAMAC is designed specifically for low-precision (2, 4, and 8-bit) quantized neural networks, 
rendering it unsuitable for general computing tasks like GEMV.
BRAMAC did not report the system-level frequency which is why we could not plot its execution time.

SPAR-2 has the longest latency across all precisions, due to its slow NEWS accumulation network,
with latency increasing almost linearly with matrix dimension.
CCB and CoMeFa-based GEMV engines have the shortest cycle latency among bit-serial architectures across all precisions.
This is due to their fast reduction algorithm based on a popcount-based adder and pipelined adder tree.
The cycle latency of {\gemvName} is significantly shorter compared to SPAR-2 but longer than CCB/CoMeFa-based implementations.
However, {\gemvName} clocks at least 2\Ax\ faster than any of the other GEMV engines. 
As a result, {\gemvName} outperforms all other GEMV engines in terms of overall execution time.
This highlights the importance of the system clock speed over the cycle latency;
despite the CCB/CoMeFa GEMV engines' shorter cycle latency, their slower clock significantly degrades the execution time.

Because {\gemvName} is utilizing only 30\% of the logic resources in U55,
the remaining resources can be used to further improve its performance.
The {\gemvName}-slice4 curves in Fig.~\ref{fig:gemvlatency} shows the latency of a variant of {\gemvName} 
with a 4-bit sliced accumulation network and a PE implementing Booth's radix-4 multiplication (default is radix-2).
This latency is estimated by adjusting the analytical model of {\gemvName} assuming no effect on the clock rate.
In terms of cycle latency, it can run almost as fast as CCB/CoMeFa-based GEMV implementations, while
significantly outperforming them in execution time.

\section{Conclusions and Future Work}
\label{sec:conclusion}
Processor In/Close to Memory (PIM) architectures have become popular frameworks 
replacing classic von Neumann architectures within domain-specific machine learning accelerators.  
This paper presented a study proposing the performance and scalability goals for PIM array-based accelerators on FPGAs.
The design, implementation, and analysis of {\gemvName} 
was presented demonstrating how a PIM-array accelerator could achieve the BRAM Fmax as the system frequency.
A scalability study was presented showing processing capacity scaling linearly with increasing BRAM density, 
even for devices with low LUT-to-BRAM ratios.
An implementation with 64K PEs was run on Alveo U55, 
clocking faster than the Tensor Processing Unit (TPU v1-v2) and Alibaba Hanguang 800.  
This breaks the myth that FPGA overlays and fabrics must clock slower than ASIC designs.

A comparative study with state-of-the-art PIM-array accelerators was presented showing IMAGine
has \mbox{2.65\Ax\ -- 3.2\Ax} faster system frequency, and significantly 
outperforms them in execution time, establishing IMAGine as the fastest and most scalable PIM array-based GEMV engine reported to date.

Our future work includes the completion of an MLIR-based compiler framework for
hardware/software codesign and application-specific customization of
{\gemvName}-like PIM array-based accelerators.

\bibliographystyle{IEEEtran}
\IEEEtriggeratref{5}
\IEEEtriggercmd{\balance}
\bibliography{IEEEabrv,ref}

\end{document}